\documentclass[]{pasj02} 
\usepackage{natbib} 
\usepackage{url}
\usepackage{comment}

\usepackage[normalem]{ulem}

\jyear{2026}
\Received{}
\Accepted{}

\begin{document} 

\title{Multi-Epoch Monitoring of the Faint Northern Component in 3C~84 during 2016--2021}

\author{
 Kotaro \textsc{Niinuma},\altaffilmark{1,2}\altemailmark\orcid{0000-0002-8169-3579} \email{niinuma@yamaguchi-u.ac.jp} 
 Motoki \textsc{Kino},\altaffilmark{3,4}\orcid{0000-0002-2709-7338}
 Nozomu \textsc{Kawakatsu},\altaffilmark{5}\orcid{0000-0003-2535-5513}
 Hiroshi \textsc{Nagai},\altaffilmark{4,6}\orcid{0000-0003-0292-3645}
 Gabriele \textsc{Giovannini},\altaffilmark{7,8}\orcid{0000-0003-4916-6362}
 Filippo \textsc{D'Ammando},\altaffilmark{7}\orcid{0000-0001-7618-7527}
 Monica \textsc{Orienti},\altaffilmark{7}\orcid{0000-0003-4470-7094}
 Kazuhiro \textsc{Hada},\altaffilmark{9}\orcid{0000-0001-6906-772X}
 Kiyoaki \textsc{Wajima},\altaffilmark{10,11}\orcid{0000-0003-3823-7954}
 and 
 Marcello \textsc{Giroletti},\altaffilmark{7}\orcid{0000-0002-8657-8852}
}

\altaffiltext{1}{Graduate school of Sciences and Technology for Innovation, Yamaguchi University, Yoshida 1677-1, Yamaguchi, Yamaguchi 753-8512, Japan}
\altaffiltext{2}{The Research Institute for Time Studies, Yamaguchi University, Yoshida 1677-1, Yamaguchi, Yamaguchi 753-8511, Japan}
\altaffiltext{3}{Division of Liberal Arts, Kogakuin University of Technology \& Engineering, 2665-1 Nakano-machi, Hachioji, Tokyo 192-0015, Japan}
\altaffiltext{4}{National Astronomical Observatory of Japan, 2-21-1 Osawa, Mitaka, Tokyo, 181-8588, Japan}
\altaffiltext{5}{National Institute of Technology, Kure College, 2-2-11, Agaminami, Kure, Hiroshima 737-8506, Japan}
\altaffiltext{6}{Department of Astronomical Science, The Graduate University for Advanced Studies, SOKENDAI, 2-21-1 Osawa, Mitaka, Tokyo 181-8588, Japan}
\altaffiltext{7}{INAF---Istituto di Radioastronomia, via Gobetti 101, I-40129 Bologna, Italy}
\altaffiltext{8}{Dipartimento di Fisica e Astronomia, Universit\'{a} di Bologna, Via Gobetti 93/2, I-40129 Bologna, Italy}
\altaffiltext{9}{Graduate School of Science, Nagoya City University, Yamanohata 1, Mizuho-cho, Mizuho-ku, 467-8501, Nagoya, Aichi, Japan}
\altaffiltext{10}{Korea Astronomy and Space Science Institute, 776 Daedeokdae-ro, Yuseong-gu, Daejeon 34055, Republic of Korea}
\altaffiltext{11}{Department of Astronomy and Space Science, University of Science and Technology, 217 Gajeong-ro, Yuseong-gu, Daejeon 34113, Republic of Korea}




\maketitle

\begin{abstract}
We present the results of multi-epoch very long baseline interferometry (VLBI) observations of the sub-parsec-scale northern jet component of 3C~84 (known as the N1 component) at 7~mm from 2016 to 2021. 
%
%
The peak position of the N1 component exhibits positional variations of approximately 0.2 mas at a distance of about 2.5~mas from the central radio core (C1) throughout the observation period.
Although a linear fit yields a small proper motion of $\mu_{\rm Dec}=0.054\pm0.019$ mas yr$^{-1}$, corresponding to an apparent velocity of $\beta_{\rm app}=0.062\pm0.021$, this displacement is comparable to the observed positional scatter, and the hypothesis that N1 exhibits systematic motion is not statistically supported. 
%
%
%
%
%
%
In contrast, the peak flux density varies by approximately a factor of two on timescales of several months. 
We propose two possible explanations for the observed behavior of N1. 
One is that N1 is the counterpart of the southern hotspot C3, but its motion has been halted by a collision with a compact and dense cloud along its path; in this case, the flux variability may be caused by changes in the free-free absorption opacity of the non-uniform cloud. 
The other possibility is that N1 is part of the northern jet that is heavily obscured by foreground free-free absorption and is observed through a gap in the absorbing medium; in this case, the variability may reflect intrinsic changes in the jet and its interaction with a non-uniform ambient medium. 
These results provide new constraints on the physical origin of N1 and on the circumnuclear environment of 3C 84.
\end{abstract}


\section{Introduction}

Relativistic jets launched from active galactic nuclei (AGNs) can transport energy and momentum from the vicinity of the central supermassive black hole into their surrounding environment, and are an important channel of AGN feedback \citep[e.g.,][]{fabian12}. At the same time, the propagation and morphology of jets are affected by the physical conditions of the surrounding medium. Very long baseline interferometry (VLBI) provides a unique means of resolving the parsec- and sub-parsec-scale structures of these jets, where their propagation and interaction with the circumnuclear environment can be investigated \citep[e.g.,][]{zensus97}. 
Multi-epoch VLBI observations can further trace the motion and evolution of individual jet components, providing important constraints on their kinematics and on the medium through which the jets propagate.

The compact radio source 3C~84, hosted by the
Seyfert galaxy NGC 1275, is located
at the center of the Perseus cluster (z=0.0176).
Thanks to its proximity, 3C~84 is one of the best laboratory to explore the central parsec-scale region of AGN
in detail.
The C3 component (see Figure~\ref{fig:sequence} for component designations), which is located about one parsec south of the central
radio core (the so-called C1 component), is a newborn approaching hotspot 
associated with the radio outburst that started in 2005
\citep[][and the references therein]{abdo09,nagai10}.
Recent VLBI observations of C3 have revealed that the jet trajectory is
disturbed by the interaction between the jet and the surrounding material
\citep[][]{nagai17,kino18,kino21}. 
The behavior of the C3 component revealed by these observations is consistent with results from a study employing the regularized maximum likelihood (RML) imaging method to achieve super-resolution \citep{foschi25}. 
These observations suggest that the ambient matter
in the central parsec-scale region of 3C~84 is 
highly non-uniform.

Regarding the northern counter-lobe component, it is generally
considered to be dimmed by free-free absorption (FFA)
in the ionized gas associated with the accretion disk \citep[][]{walker00}.
Although the counterpart of C3 was undetectable for about 10 years 
between 2005 and 2015, a new component named N1 was detected in December 2015 \citep{fujita17, wajima20}.
\citet{foschi25} also report the detection of diffuse emission to the north of C1, which is consistent with the N1 component. 
At the time of its discovery at the end of 2015, N1 was naturally interpreted as the counter hotspot propagating at approximately the same speed as C3, 
since their distances from C1 are comparable.

We focus on 43 GHz VLBI observations for following reasons. First, the high angular resolution and imaging capability at 43 GHz enable us to resolve the sub-parsec-scale structure of 3C 84 and to isolate the relatively faint N1 component from the surrounding emission. Moreover, the 43 GHz VLBI arrays provide relatively stable and well-established observing capabilities over a wide range of observing epochs, which is important for long-term monitoring of the faint and diffuse N1 component. 
Second, FFA is less severe at 43 GHz than at lower frequencies, allowing us to probe the northern side of the source that is strongly obscured at lower frequencies. At the same time, the absorption is not necessarily negligible at 43 GHz, making this frequency particularly useful for investigating the interplay between the intrinsic jet structure and the foreground absorbing medium. 
Thus, 43 GHz offers a useful balance between angular resolution, and reduced FFA, enabling us to investigate both the kinematics of N1 and its possible connection to the absorbing circumnuclear medium.

%
In this paper, we report the result of multi-epoch 43 GHz VLBI observations of the N1 component from 2016 to 2021, combining KaVA monitoring observations with archival VLBA data. We investigate the temporal evolution of the position and flux density of N1 to determine whether it exhibits systematic motion and to constrain its physical origin. 
Throughout this paper, we adopt the cosmological parameters $H_0=69.3~\mathrm{km s^{-1}~Mpc^{-1}}$ and $\Omega_\mathrm{\Lambda}=0.712$ \citep{bennett14}.
Based on these parameters, 1 milliarcsecond (mas) at the distance of 3C~84 corresponds to a linear scale of 0.35~pc. The distance to 3C~84 is $D=72$~Mpc.

\section{Observations and data reduction}
\subsection{KVN and VERA Array (KaVA) observation}
Between 2016 August 16 and 2017 December 20, we performed monthly VLBI monitoring at 43~GHz to investigate temporal variations of the pc-scale jet structure of 3C~84 with a joint mm-VLBI array in the East Asian region called KaVA \citep[][]{niinuma14a}.
KaVA consists of the Korean VLBI Network \citep[KVN, three 21 m radio telescopes;][]{lee14} and the VLBI Exploration of Radio Astrometry \citep[VERA, four 20 m radio telescopes;][]{kobayashi03} Array operated by the Korea Astronomy and Space Science Institute (KASI) and the National Astronomical Observatory of Japan (NAOJ).
%
Three additional KaVA observations were conducted in 2018 March, May, and November.
Each observation of 3C~84 was conducted with an on-source time of 5.9 hr.
All data were recorded at 1 Gbps (a total bandwidth of 256~MHz with 8 or 16 subbands), with left-hand circular polarization (LHCP), and were correlated by the Daejeon Hardware Correlator \citep{lee15a} in the Korea--Japan Correlation Center (KJCC).


\subsection{Data reduction}
Amplitude and phase calibration of the correlated visibilities were carried out using the NRAO Astronomical Image Processing System (AIPS) software \citep{greisen03} in the standard manner, and an additional amplitude correction factor was applied using the AIPS task APCAL \citep{lee15b}.
The calibrated visibilities were imported into the Caltech Difmap software \citep{shepherd97}, and imaging was performed using the standard CLEAN and self-calibration procedures.
%
However, since the N1 component is fainter than both the radio core and C3 components, we did not use clean boxes around the N1 component prior to amplitude self-calibration to avoid introducing artificial structures.
A short solution interval of 0.5--1 min was adopted because of the high S/N.%
The typical angular resolution was 0.6~mas.

\subsection{VLBA archival data}
\begin{figure}[h]
\begin{center}
    \includegraphics[width=\linewidth]{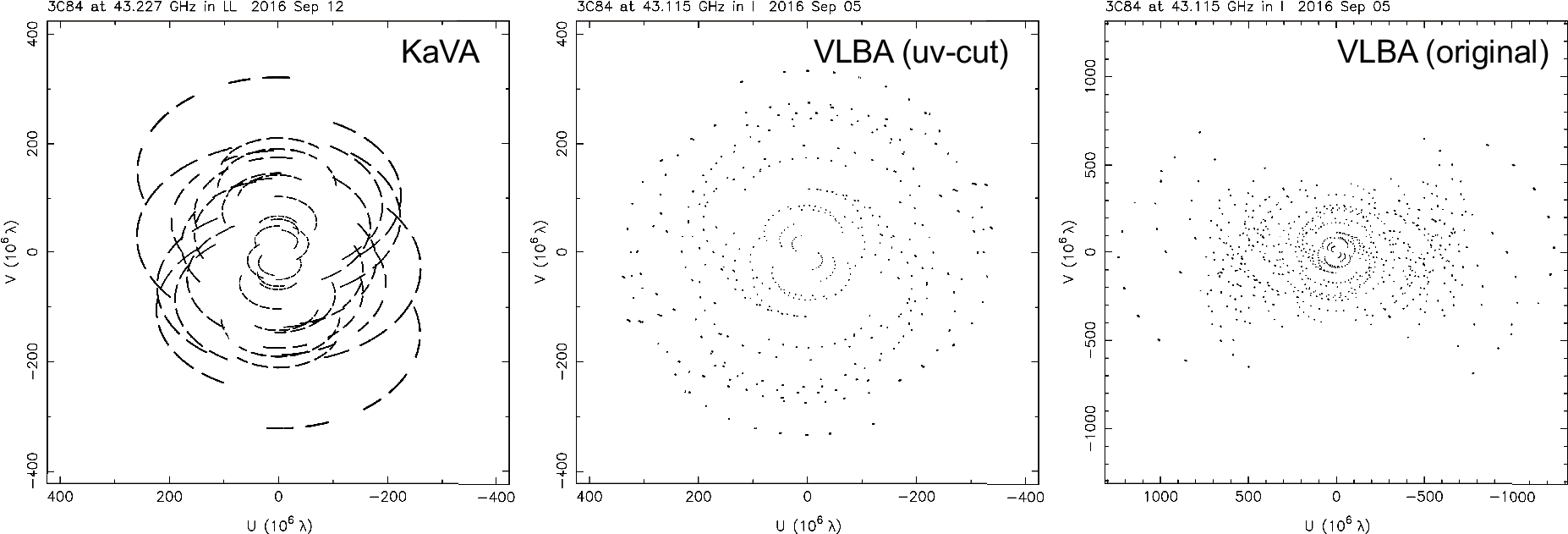}
\end{center}
\vspace{-0.1in}
\caption{
Examples of $uv$--coverage for observations of 3C~84.
Left: KaVA.
Center: VLBA, restricted to spatial frequencies up to 350~M$\lambda$.
Right: full VLBA $uv$--coverage.
The VLBA data are archival data from the blazar monitoring program led by the Boston University group.
}\label{fig:uv}
\end{figure}

To verify the KaVA results and extend the temporal coverage, we additionally analyzed archival 43 GHz VLBA datasets\footnote{\url{https://www.bu.edu/blazars/VLBA_GLAST/0316.html}}. 
We used the publicly available calibrated visibility data and CLEAN images,  and applied the same model-fitting procedure described as below.
%
To enable a direct comparison with the KaVA data, the archival VLBA visibility data were restricted to projected baseline lengths shorter than 350 M$\lambda$, corresponding to the maximum spatial frequency sampled by the KaVA observations. The filtered visibility data were then imaged using a circular restoring beam with a FWHM of 0.6 mas, identical to that adopted for the KaVA images. 
Examples of the uv-coverage before and after the baseline restriction are shown in Figure~\ref{fig:uv}.
%
All subsequent analyses were performed using these uniformly restored images.
After excluding VLBA observations with a $5\sigma$ image noise level exceeding 30~mJy~beam$^{-1}$ to ensure reliable measurements of N1, 
38 archival VLBA epochs were retained for analysis. Combined with the 18 KaVA observations, a total of 56 observing epochs were used in this study.
%
%

\begin{figure*}[htbp]
\begin{center}
    \includegraphics[width=0.85\linewidth]{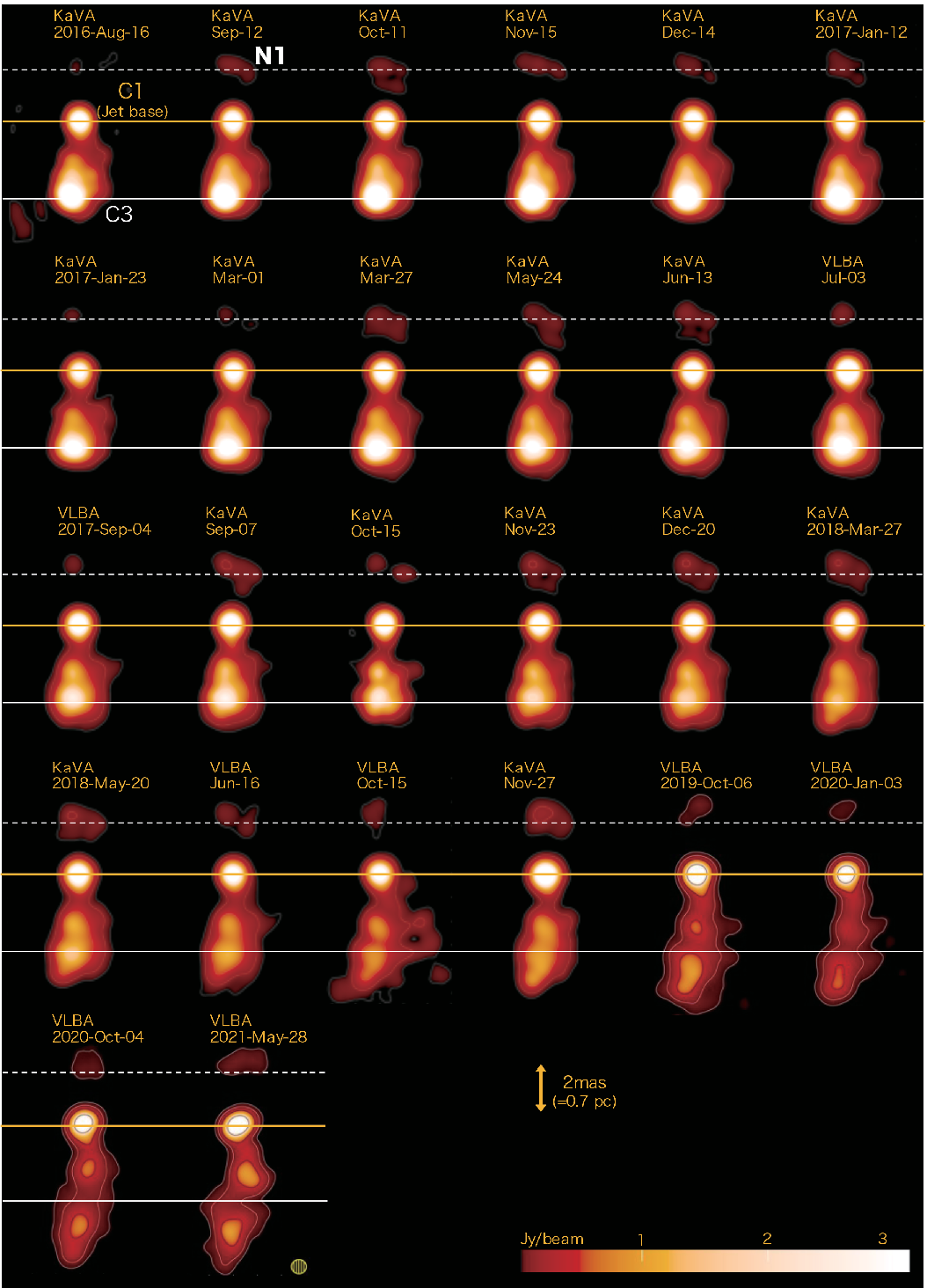}
\end{center}
\vspace{-0.1in}
\caption{KaVA and VLBA images of the sub-parsec-scale structure of 3C~84 from 2016 August 16 to 2021 May 28. All images were restored with a circular Gaussian beam of 0.6~mas, indicated by the yellow filled circle in the bottom corner of the panel for 2021 May 28. This represents the typical angular resolution achieved by KaVA observations at 43~GHz and corresponds to a linear scale of 0.21 pc.
The intensity range is shown by the color scale in the bottom-right corner with contours at levels of 30~mJy~beam$^{-1}\times$ (1, 4, 16, 64, 256).
The horizontal yellow solid, white dashed, and white solid lines indicate the positions of the brightness centroid of C1 at all epochs, N1 in 2015 December when it was first discovered \citep[][]{fujita17}, and C3 on 2016 August 16, respectively.
The positions of N1 and C3 are measured relative to the brightness centroid of C1 at the same epoch.
}\label{fig:sequence}
\end{figure*}


\subsection{Modelfit procedure}\label{mdfit}
%
%
The N1 component exhibits an extended structure elongated in the east--west direction, 
in contrast to the compact C1 and C3 components. 
Owing to its relatively low surface brightness and diffuse morphology,
the internal structure of N1 cannot be robustly constrained.
%
%
%
To evaluate an appropriate representation of the N1 structure, 
we performed Monte Carlo simulations, as described in Appendix \ref{error_estimation}. 
The simulations demonstrate that fitting the N1 region with multiple Gaussian components 
introduces large uncertainties because of the limited $uv$--coverage and 
the large number of free parameters. 
In contrast, a single Gaussian representation provides significantly 
more stable estimates of the global properties of the component.
Based on these results, we characterized N1 as a single extended component throughout this study. 
Since our primary objective is to measure the representative position and 
peak intensity of the extended emission rather than to resolve its internal substructure, 
we performed image-plane Gaussian fitting using the AIPS task JMFIT. 
%
Following the Gaussian fit, 
The N1 component was regarded as detected 
when the fitted peak intensity exceeded seven times the rms image noise ($S/N > 7$). 
Applying this criterion, N1 was detected in 50 of the 56 observing epochs. 

\section{Results}\label{result}

Figure~\ref{fig:sequence} shows the time evolution of the pc-scale jet structure from 2016 August to 2021 May, 
based on the KaVA monitoring observations supplemented by archival VLBA data.

As reported by \citet{kino21}, the southern hotspot C3 has been gradually moving southward with respect to C1. During the period from 2016 September to 2017 December, C3 exhibited a “frustrated” phase, likely caused by interactions with dense ambient clouds, and then diminished. 
In contrast, the northern jet component N1, which is considered to be the counter hotspot of C3 \citep{fujita17}, 
showed no systematic motion in its global structure over the five-year period, but exhibited a non-negligible jittering motion at $\sim2.5$ mas north of C1, as discussed in this section
\footnote{Some of the VLBA images that appear in Figure \ref{fig:sequence} were previously reported by \citet{kino21}. 
Because they focused exclusively on C1 and southern structures to examine the significant time evolution of the C3 component as mentioned above, their VLBA images did not include the northern region of C1.}.
A change in the internal structure within the N1 region was also addressed by \citet{liu26}, who suggested possible precession in this component.

\subsection{The peak position of N1}
\begin{figure}[htbp]
\centering
\includegraphics[width=\linewidth]{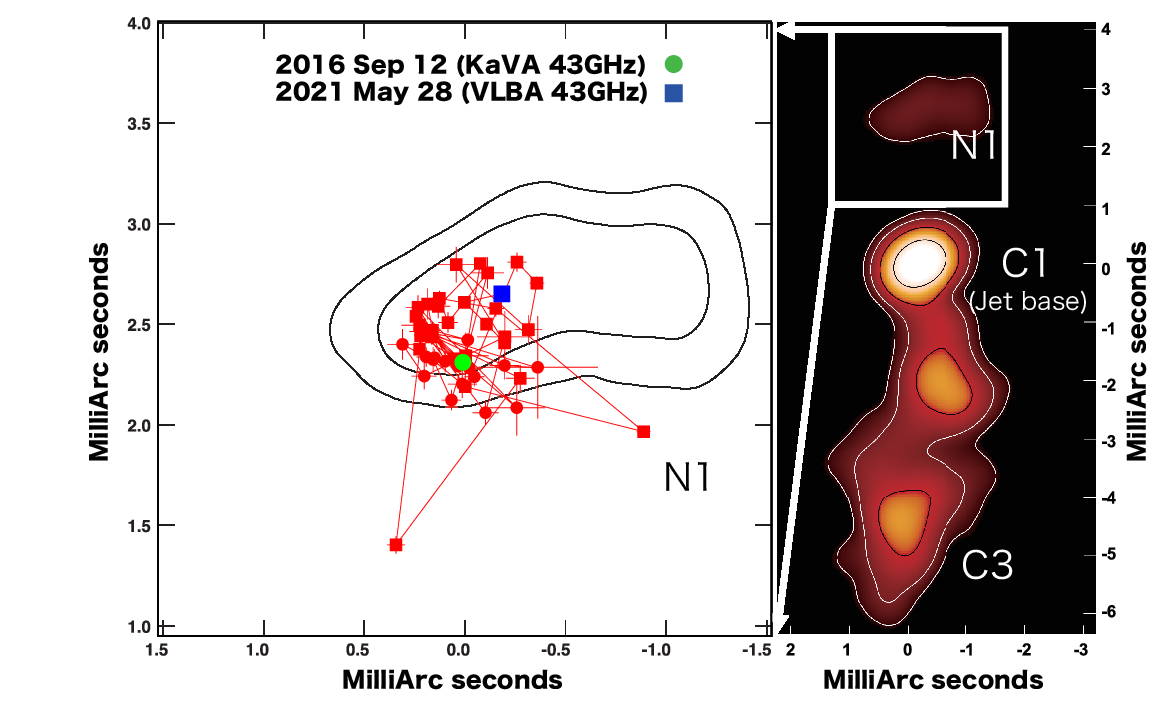}
\caption{The (\textit{left}) panel shows the location of the brightest peak of N1 at each epoch overlaid on the VLBA image obtained on 2021 May 28. 
The contours are shown at levels of 30~mJy~beam$^{-1}\times$ 1 and 2.
The green circle and the blue square indicate the first epoch (2016 September 12) and the last epoch (2021 May 28), respectively.
The (\textit{right}) panel shows the pc-scale structure of the jet base in 3C~84 observed on 2021 May 28, and is the same as that shown in Figure\ref{fig:sequence}.  
The contours start at levels of 30 mJy beam$^{-1}\times$ (1, 4, 16, 64, 256).
%
}\label{fig:n1motion}
\end{figure}

%
The changes in the location of N1's brightest peak from 2016 September 12 to 2021 May 28 are shown in Figure \ref{fig:n1motion}.
%
%
In the figure, the positional uncertainty ($\sigma_\mathrm{total}$) of the N1 component at each epoch is estimated by combining three contributions as 
$\sigma_\mathrm{total}^2 = \sigma_\mathrm{th}^2 + \sigma_\mathrm{sim}^2 + \sigma_\mathrm{str}^2$, 
where $\sigma_\mathrm{th}$ is the thermal fitting error, $\sigma_\mathrm{sim}$ is the simulated uncertainty (Appendix \ref{error_estimation}), and $\sigma_\mathrm{str}$ is an additional uncertainty associated with the finite spatial extent of the N1 structure. 
$\sigma_\mathrm{str}$ is estimated as the size of N1, modeled as a single Gaussian component, divided by the S/N. This term becomes significant when the emission is spatially extended compared with the synthesized beam. 
%
%
%
The first and last epochs are highlighted in different colors to facilitate tracking of the temporal evolution in the figure.
%
Structural changes in the radio core region in the east-west direction, as reported by e.g., \citet{paraschos22, park24}, may affect the apparent displacement of N1 because its position is measured relative to the location of C1.
However, no significant structural change is seen within 0.5~mas to the south of the core (the component labeled ``C'' in Figure 2 of \citealt{paraschos22}).
Therefore, the structural changes in the radio core region mentioned above do not significantly affect the discussion of the N1 position in the north-south direction.
%
%
%
%
To quantify the motion of N1 in declination, we derived the proper motion $\mu_\mathrm{decl}$ from the positional changes measured over a period of less than 5 years. 
We obtained $\mu_\mathrm{decl} = 0.054 \pm 0.019$ mas yr$^{-1}$, corresponding to an apparent velocity of $\beta_\mathrm{app,N1} = 0.062 \pm 0.021$. 
However, over the same period, the positional scatter of N1 is approximately 0.22 mas in right ascension and 0.24 mas in declination.
Thus, the measured displacement is comparable to the overall positional scatter of the data, and no clear systematic trend can be identified within this level of uncertainty.
These results indicate that the apparent displacement of N1 cannot be distinguished from positional jitter.
Therefore, the hypothesis that N1 exhibits systematic motion is not statistically supported.
%
%
%
%
%
%
%
%

%
In contrast, \citet[][]{fujita17} suggested that N1 may exhibit significant motion. They proposed that N1 could be the counter hotspot of C3. 
If this interpretation is correct, it would have taken about 10 years after ejection from the core for the component to appear 2 mas north of C1. 
This implies a proper motion of $\sim0.2~\mathrm{mas~yr}^{-1}$, corresponding to an apparent speed $\beta_\mathrm{app,N1}$ of $\sim0.23$. 
%
%
%
The proper motion measured during our monitoring period was less than one-third of this value, indicating that the kinematic behavior of N1 has changed substantially.

\subsection{The peak intensity of N1}
\begin{figure}[htbp]
\centering
\includegraphics[width=\linewidth]{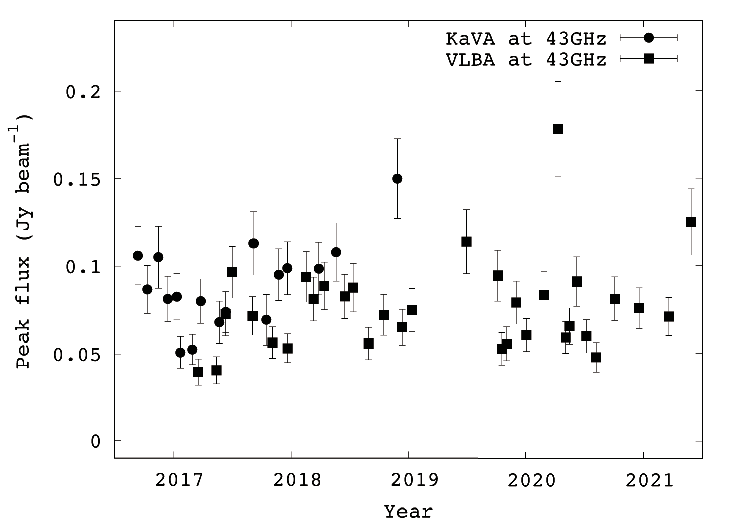}
\caption{Time variation of the peak flux density of N1, derived from a single-component fit.}\label{fig:n1lc}
\end{figure}

%
%
Figure \ref{fig:n1lc} shows the light curve of the peak flux density of N1 derived from Gaussian model fitting. The peak flux density varies by approximately a factor of two on timescales of a few months during the monitoring period from 2016 to 2018. 
The observed flux variations significantly exceed the estimated uncertainties, suggesting that they are unlikely to be dominated by measurement errors.
The error bars at each epoch represent the quadrature sum of the thermal noise error obtained from the model fitting using the AIPS task JMFIT, the simulated uncertainty described in Appendix \ref{error_estimation}, and a systematic uncertainty. 
We adopt a systematic uncertainty of 15\% in the peak flux density, as estimated from the simulations described in Appendix \ref{error_estimation}.

To quantitatively evaluate the variability, we also performed a $\chi^2$ test under the null hypothesis that the peak flux density remains constant at the weighted mean value. 
For the entire observing period, we obtained $\chi^2 = 160$ with $\nu = 49$ degrees of freedom, corresponding to a reduced chi-square of $\chi^2_\nu = 3.27$ and a p-value $P_\mathrm{no-var}(\chi^2; \nu) < 0.01$. 
This result rejects the null hypothesis, indicating statistically significant variability of N1 at a confidence level greater than 99\%. 
This conclusion is consistent with the results of the Monte Carlo simulations presented in Appendix \ref{error_estimation}, which indicate that the observed variability cannot be explained solely by the effects of the $uv$-coverage and measurement uncertainties.
%

\section{Discussion}

Our results show that N1 exhibits no systematic motion and instead displays only jittery behaviour. Here, we discuss possible explanations for its nature (e.g. a counter-hotspot of C3, a partially obscured jet, etc.). For this purpose, 
%
%
we define the south length ($L_\mathrm{south}$) as the radial distance between C1 and C3, and the north length ($L_\mathrm{north}$) as that between C1 and N1.
When the N1 component was first identified, it was suggested that N1 is the counter hotspot of C3, since $L_\mathrm{south}$ and $L_\mathrm{north}$ were nearly symmetric \citep{fujita17}.


However, Figure \ref{fig:sequence} shows that $L_\mathrm{north}$ differs from $L_\mathrm{south}$, indicating a clear asymmetry in the distances from C1.
This asymmetry is more pronounced than that reported by \citet{fujita17} and appears to increase with time.
Furthermore, N1 exhibits small-scale jitter around a position approximately 2.5~mas north of C1, while remaining nearly stationary over several years.
This raises the question of whether N1 is indeed the counter component of C3.
If N1 is the true counter hotspot of C3, the radio images suggest a scenario in which N1 has interacted with the surrounding medium, resulting in its apparent stagnation at a distance of approximately 2--3~mas north of the radio core (Case~1).
%

%
%
In another case, the counter hotspot may be fully hidden by the FFA foreground.
%
%
This scenario requires that the surrounding material on the northern side is highly non-uniform. 
%
The FFA foreground would fully obscure the counter hotspot, while the observed N1 would be a part of the jet that is partially absorbed but remains visible through a gap-window in the FFA foreground (Case~2).

Here, we discuss the physical implications of the finding that the N1 component does not exhibit statistically significant systematic motion together with its peak flux variations. 
%
The causes are summarized in Table~\ref{table:case1}.
%
%
In the case of C3, its flux density gradually decreased due to interaction with the invisible cloud \citep{kino21}. On the other hand, no systematic flux decrease in N1 can be seen over several years. This is a significant difference in the flux variability between C3 and N1.



\begin{table*}[t]
\centering
\caption{Summary of possible scenarios for N1}
\begin{tabular}{ccccc}
\hline
\hline
Cases & Motion of N1 & Diminished by & Variability by & Feasibility \\
\hline
\hline
Case 1a & Stopped by invisible cloud 
& invisible cloud's self-FFA &
invisible cloud's nonuniform FFA & possible  \\
Case 1b  & Stopped by invisible cloud  
&  FFA-foreground &
FFA-foreground & 
difficult (too fast $v_{\rm fgd}$) \\
Case 1c  & Stopped by invisible cloud 
&  FFA-foreground &
the north hotspot intrinsic
& difficult 
(too large $B$)\\
\hline
Case 2a & 
Gap window of  foreground-FFA 
&  FFA-foreground &
the north jet intrinsic 
& possible \\
Case 2b & 
Stationary feature in counter-jet
&  --- &
the north jet intrinsic
& possible \\
\hline
\end{tabular}
\\ \vspace{0.3cm}
\raggedright{\textbf{Note.}
Cases 1a, 1b and 1c represent N1 being the counterpart (counter hotspot) of C3, while Case 2a, 2b show that 
N1 is not the counterpart of C3.} 
\label{table:case1}
\end{table*}


\subsection{Case~1: N1 is the counter hotspot of C3}\label{sec:4-1}


Let us consider how N1's motion has come to halt over several years. A similar phenomenon has been reported for C3, and the potential causes have been discussed in \citet[][]{kino18,wajima20,kino21}.
Since both N1's distance from C1 and the size of its frustrated area are comparable to those of C3, it is possible that N1's motion may have been halted by a jet-cloud collision similar to that observed in C3.
The existence of other young radio lobes with asymmetric arm-lengths also suggests jet-cloud interactions \citep[][]{dallacasa13,orienti16}.
The number density required for the cloud that stopped C3's motion is $\sim10^{5-6}~{\rm cm^{-3}}$
\citep[][]{kino21}, and a similar number density is likely needed for the cloud in the case of N1.
The observed radio flux density would be
diminished and would fluctuate by FFA when the hotspot is buried in the cloud during the interaction (Case 1a in Table~\ref{table:case1}).
\citet{kino21} estimated the spatial extension of the invisible cloud by measuring the region size of the observed jittering motion of C3.
Therefore, we measured the size of the invisible cloud possibly surrounding N1 using the same strategy. 

%

\begin{figure*}[htbp]
  \begin{tabular}{cc}
    \begin{minipage}[t]{0.45\hsize}
        \centering
        \includegraphics[width=\linewidth]{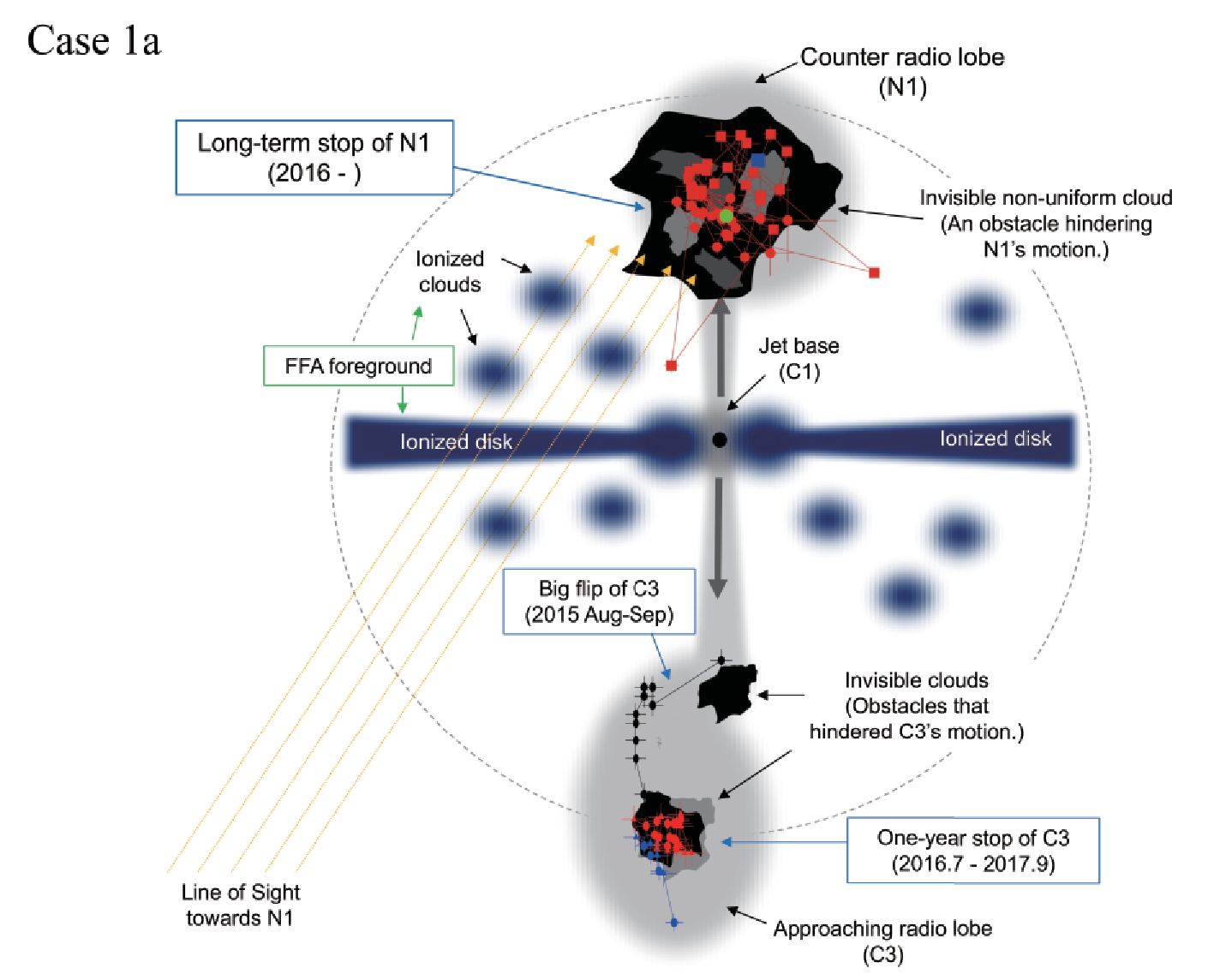}
    \end{minipage} &
    \begin{minipage}[t]{0.45\hsize}
        \centering
        \includegraphics[width=0.95\linewidth]{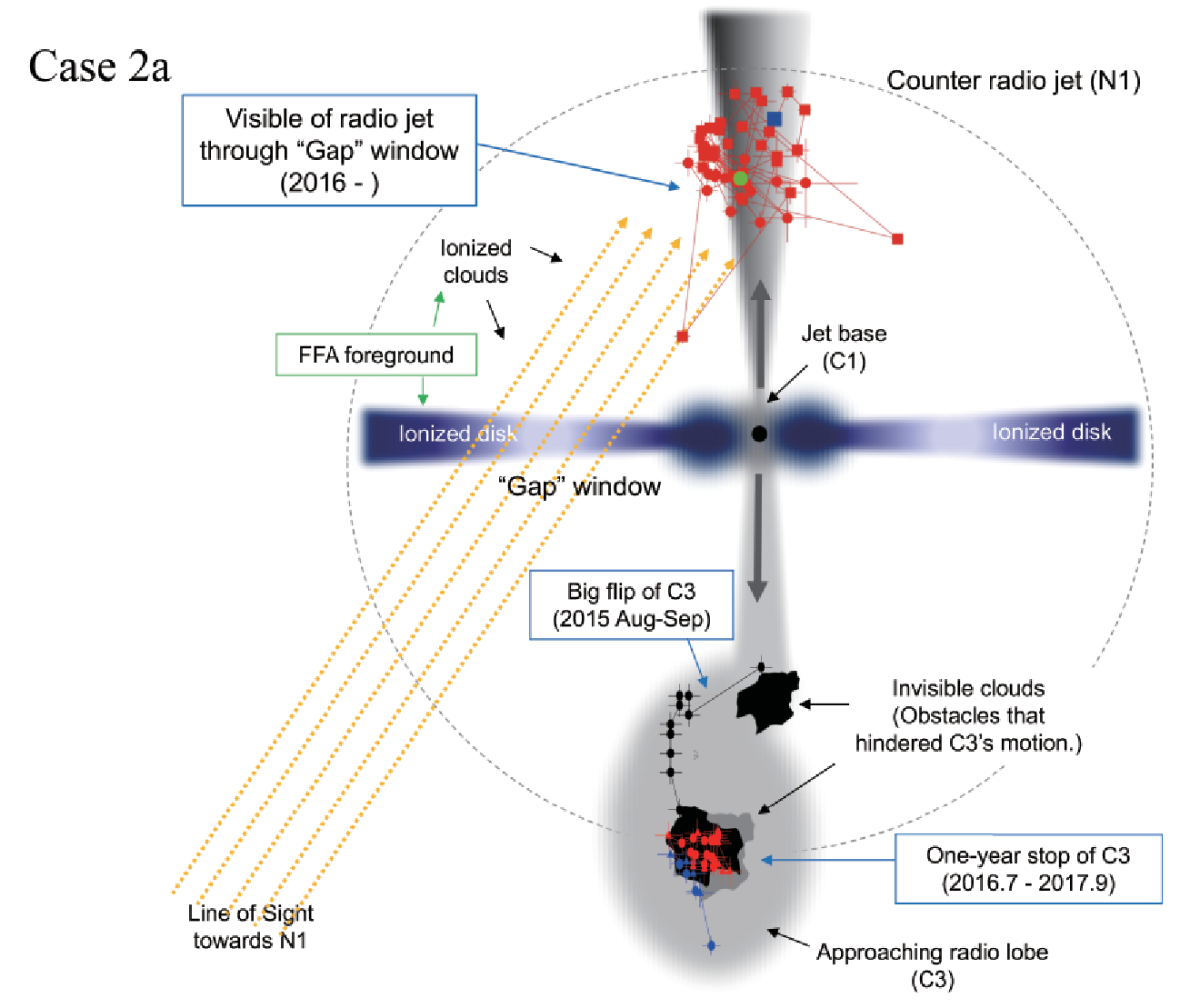}
    \end{minipage} \\
    \begin{minipage}[t]{0.45\hsize}
        \centering
        \includegraphics[width=\linewidth]{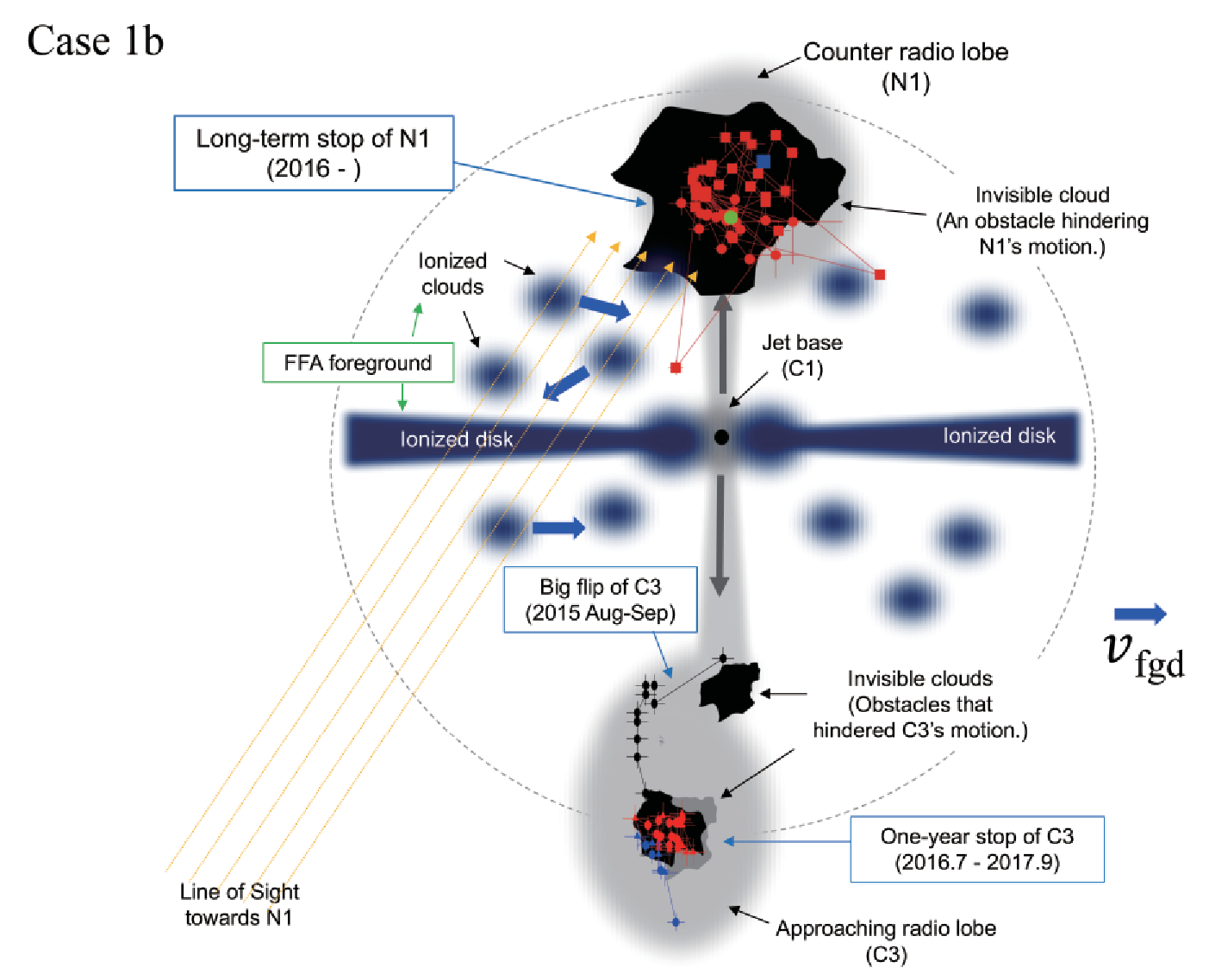}
    \end{minipage} & \vspace{-0.5in}\\
    %
    %
    \begin{minipage}[t]{0.45\hsize}
        \centering
        \includegraphics[width=\linewidth]{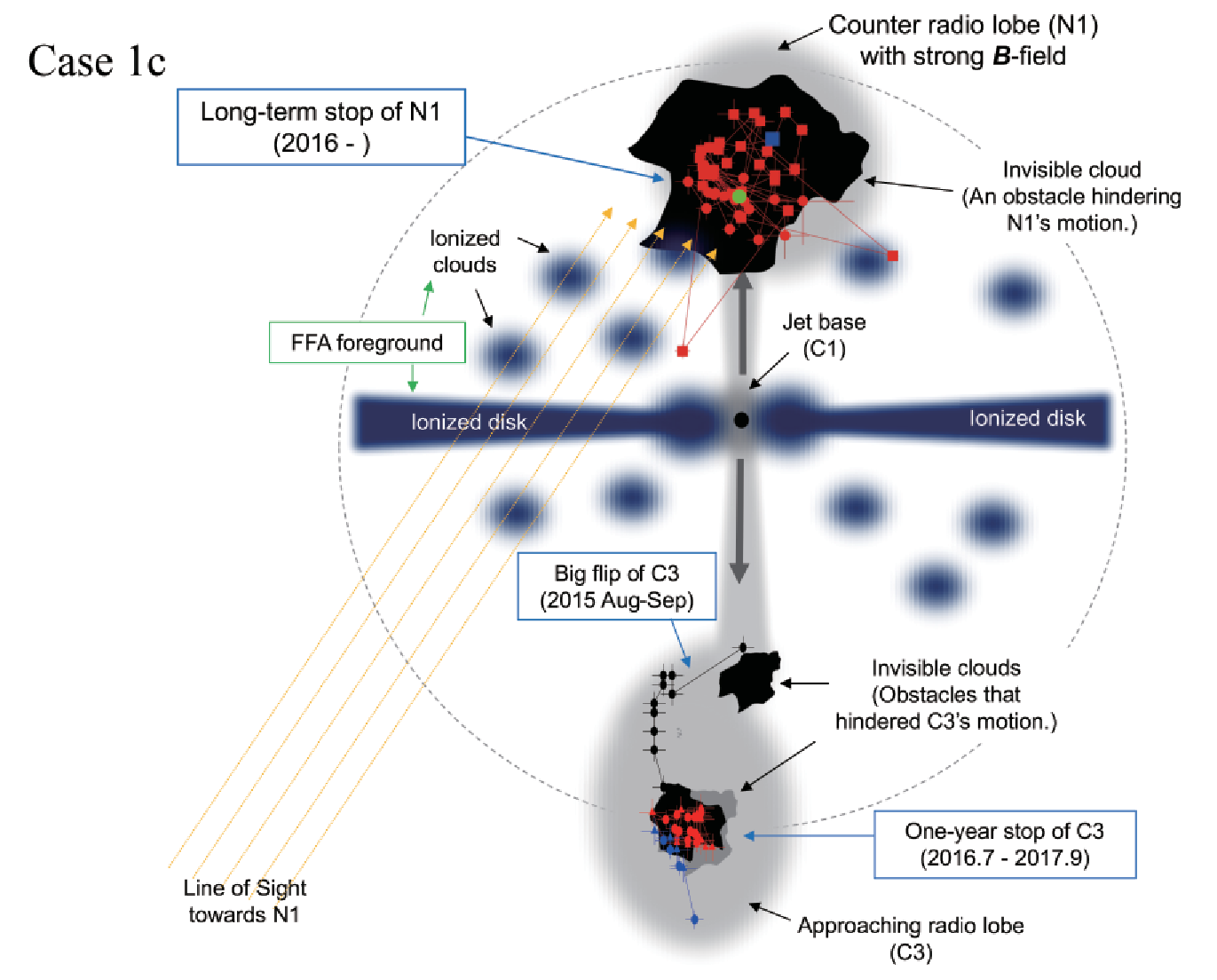}
    \end{minipage} &
    \begin{minipage}[t]{0.45\hsize}
        \centering
        \includegraphics[width=0.95\linewidth]{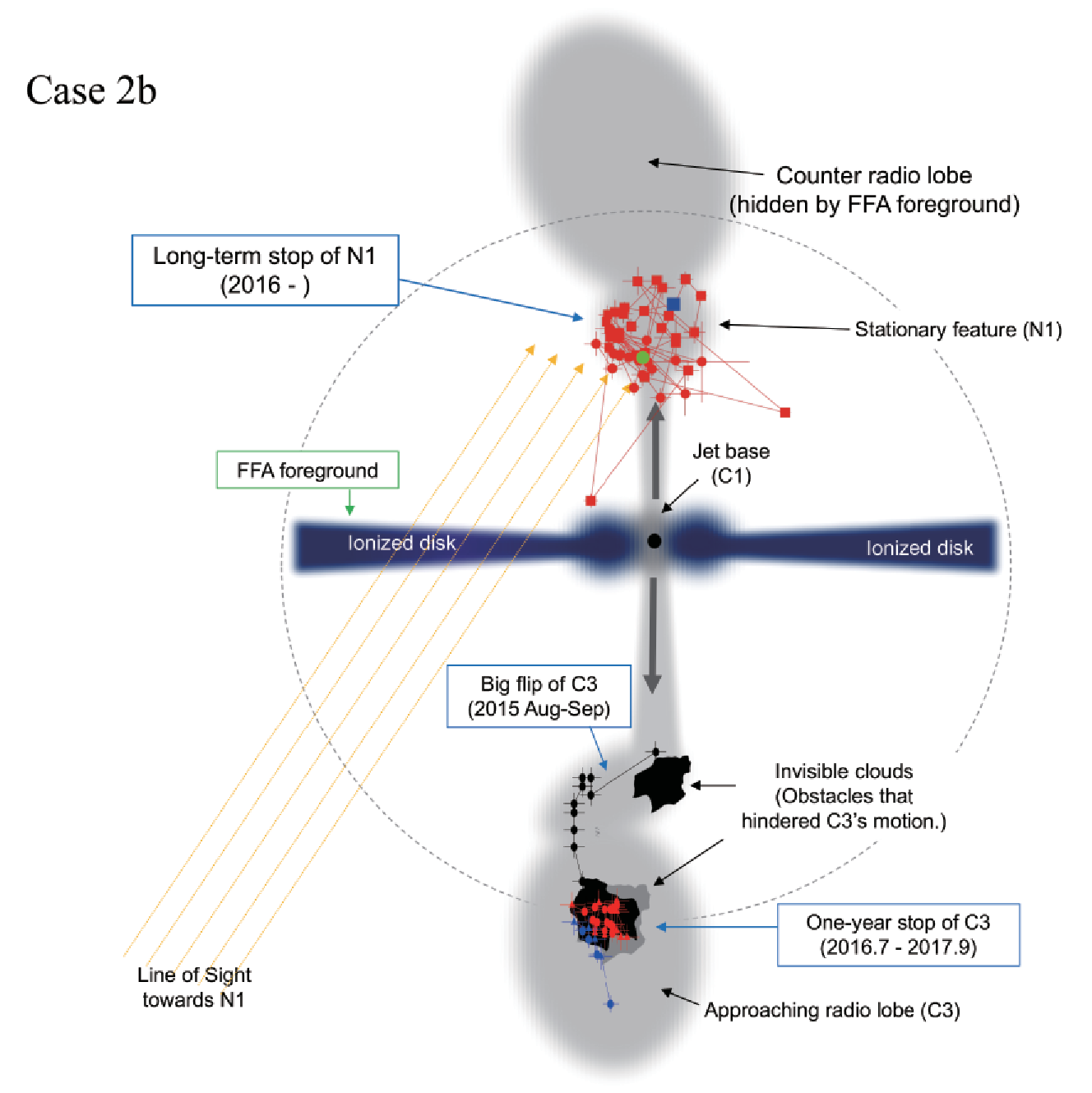}
    \end{minipage}
  \end{tabular}
    \caption{A sketch of jet--cloud interactions in 3C~84 corresponding to Cases 1a: self-FFA from the invisible cloud, 1b: fast motion of FFA-foreground, and 1c: FFA-foreground and large \textit{B}-field. 
    In these cases, an invisible cloud halts the propagation of the northern hotspot.
    The observed variability may be 
    attributed to the nonuniform density of the cloud 
    (see Table \ref{table:case1}).
    Since the jet--cloud interactions at the southern region
    were discussed by \citet{kino21}, we do not repeat them here.
    Other possible explanations for the observed phenomenon are also shown as Case 2a: FFA-foreground and the intrinsic variability of the jet and 2b: the intrinsic variability of stationary feature of the jet in the table. In these cases, the behavior of N1 is interpreted either as emission from the radio jet seen through the gap window of the FFA foreground or as a stationary feature in the counter jet.
}\label{fig:fig_summary}
\end{figure*}

Excluding the two outliers observed on 2018 August 26 and 2020 April 7, the brightness peak positions of N1 are confined within 1~mas as shown in Figure~\ref{fig:n1motion}.
In contrast, that of C3 is confined within 0.2~mas. 
This may indicate that the cloud associated with N1 is
a factor of $\sim5$ larger than that associated with C3
(see Figure~\ref{fig:fig_summary}).


Next, we discuss the possible origins of the observed flux variability of N1 shown in Figure~\ref{fig:n1lc}.
Along the line of sight,
the existence of an ionized accretion disk and clouds is suggested
\citep[e.g.,][]{walker00,fujita17,kino21}.
Therefore, the total FFA opacity can be expressed as the sum of the FFA opacity for the ionized accretion disk ($\tau_{\rm ffa,disk}$) and that for the clouds ($\tau_{\rm ffa,cloud}$),
i.e.,  $\tau_{\rm ffa}=\tau_{\rm ffa,disk}+\tau_{\rm ffa,cloud}$.
While $\tau_{\rm ffa,disk}$ is expected to be constant on a timescale of a month,
$\tau_{\rm ffa,cloud}$ may vary due to its non-uniformity.
Given the observed variability,
$\tau_{\rm ffa,cloud}\gtrsim \tau_{\rm ffa,disk}\sim O(1)$ holds, and
the increase in FFA opacity
 can be approximated as $\Delta \tau_{\rm ffa} \approx \Delta \tau_{\rm ffa,cloud}$.

Case 1a represents a scenario in which the flux variability is caused by the non-uniformity of the invisible cloud that halts the motion of the northern jet.
The FFA opacity for ionized clouds can be expressed as
\begin{align}
\frac{\Delta \tau_{\rm ffa,cloud}}{\tau_{\rm ffa,cloud}}
\approx
\frac{\Delta L_{\rm cloud}}{L_{\rm cloud}}
+2\frac{\Delta n_{\rm cloud}}{n_{\rm cloud}}
-1.5\frac{\Delta T_{\rm cloud}}{T_{\rm cloud}}
\end{align}
where $T_{\rm cloud}$, $n_{\rm cloud}$, and $L_{\rm cloud}$
are
the temperature, number density, and
the path length of the cloud, respectively
\citep[e.g.,][]{wajima20}.
As $\Delta \tau_{\rm ffa,cloud}$ increases,
the corresponding decrease in the observed flux density is given by
\begin{align}
\frac{F_{\nu}}{F_{\nu,\rm int}}=\exp(-\Delta\tau_{\rm ffa, cloud}),
\end{align}
where $F_{\nu,\rm int}$ is the intrinsic flux density of N1.
From Figure~\ref{fig:n1lc}, it is found that
$0.5 \lesssim F_{\nu}/F_{\nu,\rm int}\le 1$.
Accordingly, 
we obtain
$0\le \Delta \tau_{\rm ffa,cloud} \lesssim 0.7$.

%

For simplicity, we assume $\tau_{\rm ffa,cloud} = 1$ as a representative case for an order-of-magnitude estimate.
Then, the observed variability can be explained if 
$\Delta L_{\rm cloud}/L_{\rm cloud}$, 
$2\Delta n_{\rm cloud}/n_{\rm cloud}$, or 
$1.5\Delta T_{\rm cloud}/T_{\rm cloud}$ 
varies by up to $\sim 0.7$. 
Therefore, it is plausible that the observed flux variability is caused by the non-uniformity of the invisible cloud.

Interestingly, \citet{morganti23} indicates that
H I~absorption is visible at a slightly larger scale of a few tens of pc from SMBH, but is absent around the radio core and the pc-scale jet region. 
Therefore, there seems to be no neutral hydrogen atomic gas inside a region of $\sim 1$~pc. 
On the other hand, 
\citet{oosterloo24} reports the existence of radially infalling cold molecular gas toward the central region of AGN in the $\sim100$ pc region from the SMBH. 
%
\citet{nagai19} implies the presence of a molecular cloud in the central $\sim1$ pc region, based on observations of HCN(3-2) and HCO\(^+\)(3-2) absorption lines.
Regardless, as mentioned above, a gas with a high number density is required to halt the motion of N1. 
%
Further observations are needed to determine whether the entity interacting with N1 is molecular (neutral) gas or ionized gas.

Case 1b represents the scenario in which the flux variability is caused by motion of the FFA foreground cloud\footnote{Actually, there is the example of PKS1718-649 that indeed suggests the possibility of variability in the foreground over time \citep[][]{tingay15, macquart16}.}. 
We consider the possibility that this cloud is located near N1. 
%
%
If the angular size of the cloud ($\theta_{\rm cloud}$) is much larger than that of N1 ($\theta_{\rm N1}$), the observed variability of N1 would be difficult to account for.
Conversely, if $\theta_{\rm cloud} \ll \theta_{\rm N1}$, the cloud is too small to obscure N1. 
Therefore, we will focus on the case 
$\theta_{\rm cloud}\approx \theta_{\rm N1}$.
%
%
Using the observed variability timescale $t_{\rm var}$ and the linear scale per mas $D$, the cloud velocity is estimated by $v_{\rm cloud} \approx D\,\theta_{\rm cloud}/t_{\rm var}$. Plugging in typical values yields:
\begin{align}
   \frac{v_{\rm cloud}}{c} \approx 
      1.0 \,\biggl(\frac{D}{0.35\ {\rm pc}}\biggr)\,
      \biggl(\frac{\theta_{\rm cloud}}{0.1\ {\rm mas}}\biggr)\,
      \biggl(\frac{t_{\rm var}}{30\ {\rm days}}\biggr)^{-1}.
\end{align}
Such a relativistic velocity for the cloud is unrealistic. Therefore, Case 1b is unlikely to explain N1’s variability.

Case 1c considers the sceanario in which the variability of N1 is caused by synchrotron cooling in N1.
The observed variability is on the order of
$t_{\rm var}\sim 10^{2}~{\rm days~(a~few~months)}$.
Conversely,
the synchrotron cooling timescale ($t_{\rm syn}$) is given by 
\begin{align}
t_{\rm syn}\approx 50 \,\biggl(\frac{B_\mathrm{N1}}{1~{\rm gauss}}\biggr)^{-3/2}\,
\biggl(\frac{\nu}{43~{\rm GHz}}\biggr)^{-1/2}~{\rm days}~.
\end{align}
Where $B_\mathrm{N1}$ is the average magnetic field strength in N1.
Since $B_\mathrm{N1}$ should be significantly smaller than the magnetic field at the jet base \citep[taken to be $\sim 10^{-2}$ gauss; e.g.,][]{oh22},
it is clear that $t_{\rm syn} > t_{\rm var}$ holds at N1.
Therefore, synchrotron cooling will not be effective,
and we conclude that Case 1c is not realistic.

In summary, Case 1b requires an unrealistically high gas velocity, while Case 1c requires an unrealistically strong magnetic field to reproduce the observed flux variability. Therefore, Case 1a appears to be the most feasible scenario among the Case 1 scenarios.

 \subsection{Case~2: N1 is NOT the counter hotspot of C3}\label{sec:4-2}

 In case 2, we discuss a possible origin of the flux variability
 when N1 is not the counter hotspot of C3.
Instead, N1 is interpreted as part of the northern jet that is visible through a foreground region of smaller $\tau_{\rm FFA}$.
A region of such low $\tau_{\rm FFA}$ is indeed suggested by RadioAstron observations of the counter-jet \citep{giovannini18}.

In contrast to the scenario involving interaction between the northern-jet tip and the surrounding material, changes in the intrinsic jet power or the position angle of the northern jet could also trigger the observed variability.
%
The flux variability may occur when the northern jet interacts with non-uniform structures in the ambient medium. Such interactions can be triggered by a change in jet direction (possibly due to precession)
\citep[e.g.,][]{hiura18,zhang22,kam24}.
Fluctuations in synchrotron emissivity caused by instabilities of the boundary shear layer between the jet and the ambient medium \citep[e.g.,][]{rieger04,sironi21} may also play a role. 
%
%
%
%

Another possibility is that the N1 component is a stationary feature in the counter-jet. 
In this case, the collision of the newly ejected jet component with the stationary feature could lead to a variation of N1's flux. 
The timescale of this variation is expected to range from several months to a year \citep[e.g.,][]{doi18}. 
%
%
%
In these cases, the true hotspot may lie further north. However, it is not detected because the image sensitivity of our data is limited, especially due to the strongly varying foreground FFA inhomogeneities (see \S\ref{sec:4-1}).


Thus, both Case 2a and Case 2b can potentially explain the observed stationarity or jittering motion, and flux variability, and thus both represent feasible scenarios.

\subsection{Alternative possibility in surrounding environment of N1}

%
We discussed two possibilities. In Case 1, N1 is the counter-hotspot of C3. In Case 2, N1 is instead part of the northern jet that is seen through a lower-opacity region (gap-window) of the foreground absorber, or N1 is a stationary feature in the counter-jet.

Other possible causes of the asymmetry include Doppler boosting or the absence of a counter-hotspot.
%
%
First, we check whether Doppler boosting can explain the slow motion of N1.
The apparent speed is given by $\beta_\mathrm{app,N1}=\beta_\mathrm{N1}\sin\theta/(1-\beta_\mathrm{N1}\cos\theta)$.
Here,  
$\beta_\mathrm{N1}$ is the intrinsic velocity of N1 (in the units of $c$), 
and 
$\theta$ is the jet viewing angle. 
C3 exhibited southward motion with a proper motion of $>0.5$ mas yr$^{-1}$ before and after the 2016.7--2018.0 `frustrated phase' \citep{kino18,kino21}.
%
%
%
Assuming $\theta=65^{\circ}$, as discussed by \citet[][]{fujita17}, and that the intrinsic velocity of the approaching jet ($\beta_\mathrm{C3}$), estimated at approximately 0.50, is equivalent to $\beta_\mathrm{N1}$, we expect $\beta_\mathrm{app,N1}$ to be roughly $82\%$ of the apparent speed of C3 ($\beta_\mathrm{app,C3}$).
%
%
On the other hand, measured proper motion of N1 is only $0.054\pm0.019$ mas yr$^{-1}$ (i.e. $\lesssim10\%$ of $\beta_\mathrm{app,C3}$). 
Consequently, it is difficult to explain N1's slow motion by Doppler boosting.

%

Second, if the surrounding material on the northern side is denser or more irregular than that on the southern side, it could prevent the formation of the counter-hotspot.
This idea has been suggested by theoretical works
\citep[e.g.,][and references therein]{DeYoung93,kawakatu09}.
%
Similar cases occur in hybrid morphology radio sources (HYMORS), where one of the radio lobes lacks a terminal hotspot \citep[e.g.,][]{gopal-krishna00,giroletti03,gawronski06}.
However, in the case of 3C~84, the rather symmetrical north-south radio lobe was formed by the 1959 outburst \citep{nesterov95}.
Therefore, if this scenario were true,
significant changes in the distribution of the surrounding material would have to occur afterward.


The region extending from C1 to about 2 mas north in VLBI images of the 2010s has been obscured by FFA \citep[e.g.,][]{fujita17, wajima20}.
By contrast, the same region was visible at frequencies above 15~GHz in 1995 \citep[Figure 1 in][]{walker00}.
This suggests that the FFA foreground was denser in the 2010s, causing stronger FFA on the northern side, unless the surface brightness of the northern jet varied significantly.
This scenario also requires changes in the distribution of
the surrounding material on a timescale of about 10 years.

While an asymmetric environment could qualitatively account for the apparent asymmetry between the two sides, the requirement that the ambient density on the northern side (i.e., the N1 side) be substantially higher than that on the southern side is inconsistent with the recent constraints of \citet{kawakatu25}. Furthermore, producing such a large density contrast between the north and south would require a significant misalignment between the jet axis and the rotation axis of the circumnuclear disk (CND), which acts as the foreground absorber responsible for the FFA \citep[e.g., Figure 5 of][]{chris24}. To account for the observations, such a change in the relative alignment between the jet axis and the CND rotation axis would need to occur on a timescale as short as $\sim$10 years, which is physically implausible. Therefore, we conclude that the scenario in which the apparent stationarity or jittering motion is explained solely by an asymmetric ambient density distribution is disfavored.

Taken together, the scenarios discussed in this section suggest that the apparent stationarity or jittering motion and monthly-scale flux variability of N1 are most favorably explained either by a jet-cloud collision with non-uniform FFA (Case 1a) or by emission from the northern jet seen through a low-opacity gap window or from a stationary feature in the counter-jet (Cases 2a/2b). To firmly distinguish between these scenarios, future high-frequency VLBI observations at shorter mm wavelengths (e.g., $\geq$86 GHz), which are less affected by foreground FFA, will be essential.

\section{Conclusion}

In this paper, we present the results of multi-epoch VLBI observations of the sub-parsec-scale northern jet component N1 of 3C 84 at 7 mm during 2016-2021. We find that the global structure of the northern component N1 exhibits no statistically significant systematic motion, but instead shows positional jitter around a mean position approximately 2.5 mas north of the central radio core C1. Its peak flux density also varies by approximately a factor of two on timescales of several months.
We consider two scenarios to explain these observed behaviors. In one scenario, N1 is the counterpart of the southern hotspot C3, but its motion is halted by a collision with a compact and dense cloud along its path. In this case, the observed flux variability can plausibly be produced by changes in the free-free absorption opacity of the non-uniform cloud surrounding N1. In the other scenario, N1 is part of the northern jet that is heavily obscured by foreground free-free absorption and is visible through a lower-opacity gap window in the absorbing medium. In this case, the flux variability may reflect changes in the intrinsic jet power or direction and its interaction with a non-uniform ambient medium.
These results provide new constraints on the physical origin of N1 and on the physical conditions of the parsec-scale circumnuclear environment of 3C 84.

In order to reveal the origin of N1, it is crucial to understand the physical condition of the pc-scale circumnuclear environment of 3C~84.
Shorter mm-VLBI observations (e.g., VLBI at 86~GHz or higher frequencies), which make it possible to avoid strong FFA effects from the foreground, will play a key role in further studies of the circumnuclear environment in the near future.

\begin{ack}
%
%
This work made use of the East Asian VLBI Network (EAVN), which is operated
under cooperative agreement by the National Astronomical Observatory of Japan (NAOJ),
the Korea Astronomy and Space Science Institute (KASI), the Shanghai Astronomical Observatory
(SHAO), the Xinjiang Astronomical Observatory (XAO), the Yunnan Astronomical
Observatory (YNAO), the National Astronomical Research Institute of Thailand (Public
Organization; NARIT), and the National Geographic Information Institute (NGII), with
operational support by Ibaraki University (for the operation of Hitachi 32-m and
Takahagi 32-m telescopes), Yamaguchi University (for the operation of Yamaguchi 32-m telescope), and Kagoshima University (for the operation of VERA Iriki antenna).
This work is partially supported by JSPS/MEXT KAKENHI (grants JP18K03656, JP18H03721, JP18K03709, JP21H01137, 19K03918, JP26K21725, and JP25H00660). 
K.H.is supported by Daiko Foundation (grant J0SE807004).
This work was supported in part by Grant-in-Aid for Outstanding Research Group Support Program in Nagoya City
University Grant Number 2530002.
This study makes use of VLBA data from the VLBA-BU Blazar Monitoring Program (BEAM-ME and VLBA-BU-BLAZAR;
http://www.bu.edu/blazars/BEAM-ME.html), funded by NASA through the Fermi Guest Investigator Program. The VLBA is an instrument of the National Radio Astronomy Observatory. The National Radio Astronomy Observatory is a facility of the National Science Foundation operated by Associated Universities, Inc.
\end{ack}

East Asia VLBI Network (EAVN)

%

%
%

%
\bibliographystyle{aasjournal}
\bibliography{3c84}


\appendix

\section{Monte Carlo Error Estimation of Structural Models for Faint Components}\label{error_estimation}

\begin{figure*}[htbp]
    \centering
    \includegraphics[width=\linewidth]{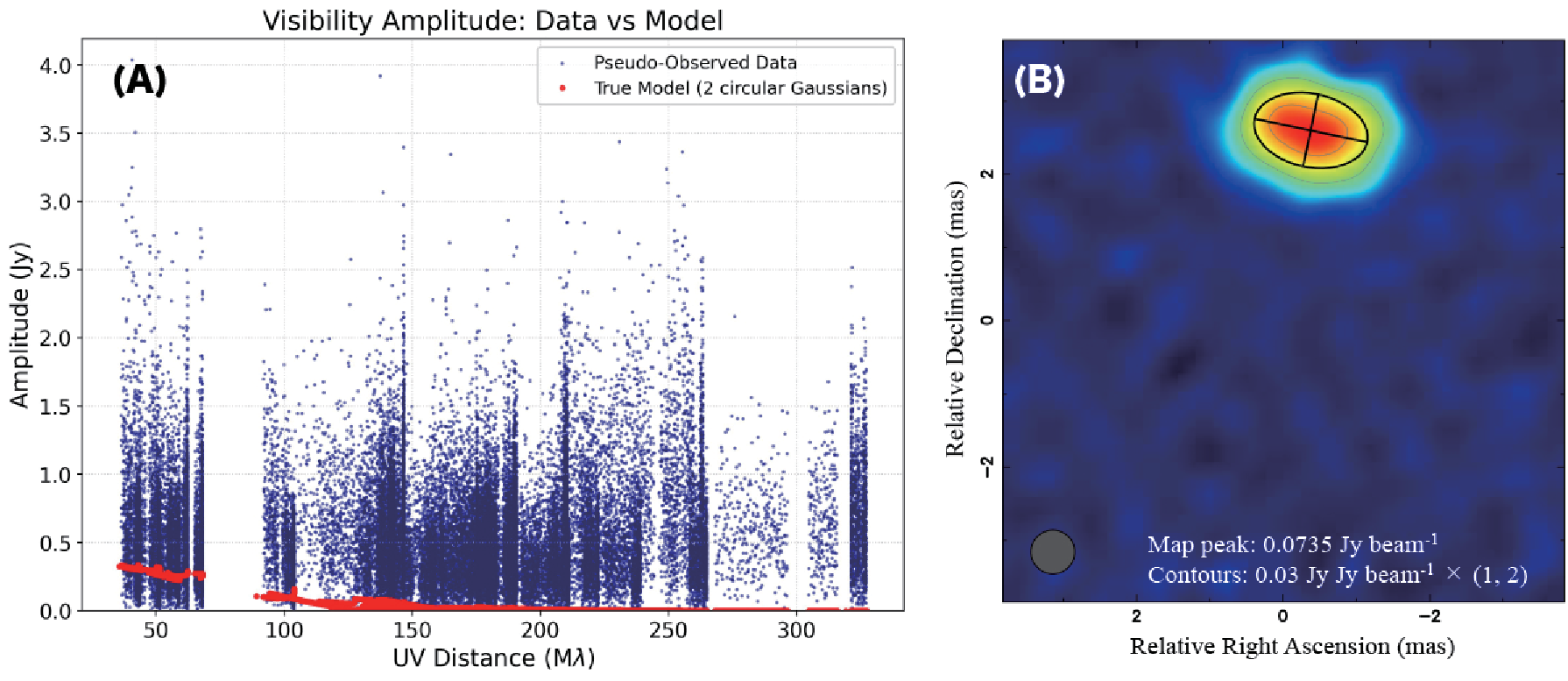}
    \caption{An example of simulated visibilities (panel A) and the corresponding reconstructed image with a model fit using a single elliptical Gaussian component (panel B). The simulated visibilities were generated using the $uv$-coverage and visibility noise level of the KaVA observation performed on 2016 October 11. {Alt text: Schematic illustrations of five possible scenarios for the environment surrounding the N1 component, as discussed in this paper.}\label{fig:example}}
\end{figure*}


To evaluate the reliability of the model fitting for the diffuse N1 component, we performed Monte Carlo simulations. Because the $uv$--coverage and the noise level of visibility vary from epoch to epoch, the simulations were carried out independently for all 50 epochs in which N1 was detected. For each epoch, the observed $uv$--coverage and visibility noise were used to reproduce the actual observing conditions. The figures presented in this appendix show one representative example to illustrate the simulation procedure and its typical results.

We first constructed an intrinsic source model consisting of two circular Gaussian components representing the N1 structure. This two-component configuration was adopted to mimic a moderately extended feature with internal substructure comparable to the synthesized beam size. 
The flux density of each component was set to 0.2~Jy. The component positions ($\Delta$RA, $\Delta$Dec) were assumed to be (0~mas, 2.7~mas) and (-0.8~mas, 2.5~mas), respectively, and the FWHM of each Gaussian component was fixed at 1.0~mas.
The adopted component sizes and their separation are approximately 10\%--50\% larger than the synthesized beam, in order to mimic a marginally resolved extended structure.

Using this intrinsic model, simulated visibilities were generated by sampling the model with the actual $uv$-coverage corresponding to each observational epoch. To mimic realistic observational conditions, white noise was generated with a standard deviation equal to the visibility uncertainties measured at each data point for the corresponding epoch, and this noise was added to the simulated visibilities to produce pseudo-observed datasets.

Model fitting was then performed for these pseudo-observed datasets in the $uv$-domain using the \textit{Difmap} task modelfit. Two fitting schemes were examined: (1) a modelfit to a single elliptical Gaussian component and (2) a modelfit to two circular Gaussian components. 

Figure \ref{fig:example} shows one representative realization of the simulated visibility data and the corresponding reconstructed image.
As illustrated in the mock-observed visibility data (Figure \ref{fig:example}, panel A), the noise level at individual visibility points is significantly larger than the model visibilities corresponding to the assumed N1 structure.
Nevertheless, in the reconstructed two-dimensional images produced from these pseudo-observed datasets, the component can still be detected with sufficient brightness relative to the image noise level. 
An example image obtained after the model fitting with a single elliptical Gaussian component is shown in Figure \ref{fig:example} (panel B).

\begin{figure*}[htbp]
    \centering
    \includegraphics[width=\linewidth]{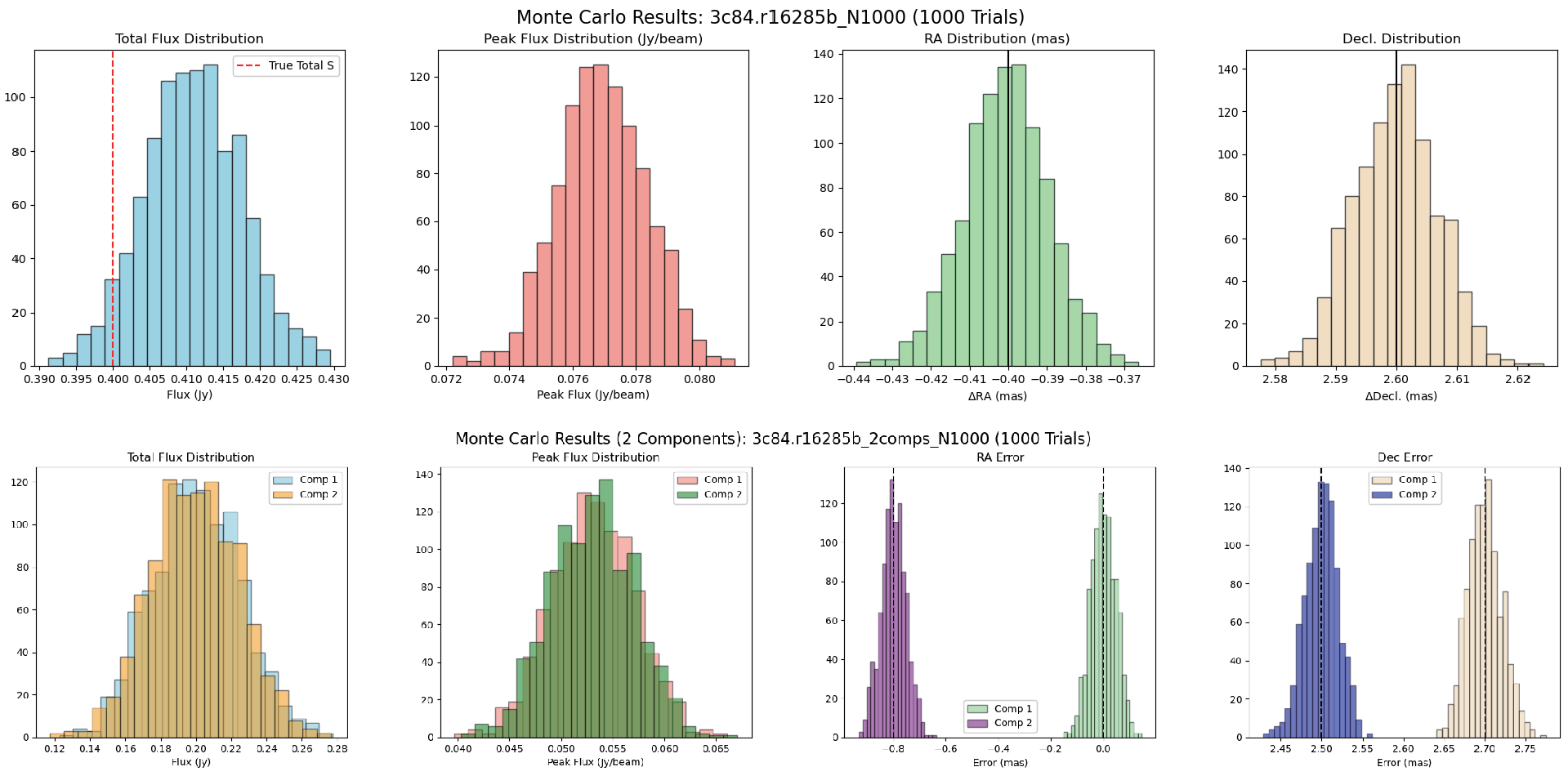}
    \caption{Representative results of the Monte Carlo simulations using the $uv$-coverage and visibility noise level of the KaVA observation conducted on 2016 October 11. 
    The intrinsic model consists of two circular Gaussian components used to generate the simulated visibilities.
    The simulations were performed with 1000 trials for two model-fitting schemes: (1) a modelfit using single elliptical Gaussian component (upper panels) and (2) a modelfit using two circular Gaussian components (lower panels).
    Each panel shows, from left to right, the distributions of the recovered total flux density, peak flux density, and component positions in right ascension and declination.
    The vertical red dashed line indicates the sum of the flux densities of the two intrinsic components. 
    The vertical black solid line shows the position of the brightness centroid of the intrinsic model, while the vertical black dotted lines indicate the positions of the individual intrinsic components. \label{fig:monte_carlo}}
\end{figure*}

For each of the 50 epochs in which N1 was detected, the above procedure was repeated 1000 times. 
%
The standard deviations of the recovered total flux densities, peak flux densities (measured using a 0.6~mas restoring beam), and component positions were calculated for each epoch (see Figure \ref{fig:monte_carlo}).

%
%

%
%
The simulation results consistently show that fitting two circular Gaussian components introduces relatively large uncertainties for all 50 epochs. 
In this case, the signal-to-noise ratio (S/N) of the fitted parameters is low, resulting in standard deviations of the peak flux densities that amount to $4$--$18\%$ of the measured values.
The positional standard deviations are $0.023$--$0.135$~mas in right ascension and $0.010$--$0.060$~mas in declination. 
In contrast, when the extended structure is treated as a single component and modeled with a single elliptical Gaussian component, the uncertainties are significantly reduced. 
%
This is because modeling the structure as a single component reduces the number of free parameters, leading to a higher S/N for the fitted parameters. 
The standard deviations are $1$--$4\%$ for the peak flux densities, and about $0.005$--$0.024$~mas in right ascension and $0.004$--$0.023$~mas in declination. 
We consider these values to represent a component of the uncertainties associated with the model fitting.
In addition, the simulations show that the peak flux density recovered from the single-component model varies between $73$ and $85$~mJy~beam$^{-1}$ over the 50 observational epochs, even though the intrinsic source structure remains unchanged. 
This corresponds to a variation of up to $\sim15\%$, which we adopt as the systematic uncertainty in the peak flux measurements in this study.
The position of the brightness centroid remains stable within $\pm0.001$~mas in both right ascension and declination. 
These results indicate that the internal structural variations within the extended N1 feature cannot be reliably constrained with the present $uv$-coverage and sensitivity. 
Therefore, it is more appropriate to treat the N1 region as a single extended structure and to discuss its global structural evolution rather than attempting to interpret variations in internal sub-components.
This supports the use of the single-component representation adopted in Section \ref{mdfit} and applied throughout Section \ref{result}. 

\end{document}